\documentclass[a4paper,twocolumn,showpacs,preprintnumbers,amsmath,amssymb,nofootinbib]{revtex4}
\input psfig.sty 
\usepackage{graphicx}

\begin{document}

\title{Time and distance constraints on accelerating cosmological models}

\author{M. A. Dantas$^{1}$\footnote{E-mail: aldinez@on.br}}

\author{J. S. Alcaniz$^1$\footnote{E-mail: alcaniz@on.br}}

\author{D. Mania$^{2,3}$\footnote{E-mail: mania@phys.ksu.edu}}

\author{Bharat Ratra$^2$\footnote{E-mail: ratra@phys.ksu.edu}}

\address{$^1$Departamento de Astronomia, Observat\'orio Nacional, 20921-400, Rio de Janeiro - RJ, Brasil}

\address{$^2$Department of Physics, Kansas State University,  116 Cardwell Hall, Manhattan 66506 USA}

\address{$^3$ Center for Elementary Particle Physics, Ilia State University, 3-5 Cholokashvili Ave., Tbilisi 0162, Georgia}

\date{\today}

\begin{abstract}
The absence of guidance from fundamental physics about the mechanism behind cosmic acceleration has given rise to a number of alternative cosmological scenarios. These are based either on modifications of
general relativistic gravitation theory on large scales or on the existence of new fields in Nature. In this paper we investigate the observational viability of some accelerating cosmological models in light of 32 age measurements of passively evolving galaxies as a function of redshift and recent estimates of the product of the cosmic microwave background acoustic scale and the baryonic acoustic oscillation peak scale. By using information-criteria model selection, we select the best-fit models and rank the alternative scenarios. We show that some of these models may provide a better fit to the data than does the current standard cosmological constant dominated ($\Lambda$CDM) model.

\end{abstract}


\maketitle

\section{Introduction}
\label{intro}

There is fairly convincing observational evidence that the Universe is now undergoing accelerated expansion. In the context of Einstein's general theory of relativity (GR) this amounts to saying that some sort of dark energy, constant or that varies slowly with time and in space, dominates the current energy budget of the cosmos. On the other hand, it is also conceivable that the current accelerated expansion could be an indication that GR does not provide an adequate description of gravity on cosmological length scales. For recent reviews, see \cite{review}.

While there are many options, if GR is assumed to be the correct model of gravitation, the current consensus opinion is that the simplest possible driver of this accelerated expansion is a cosmological constant,
$\Lambda$, a spatially homogeneous fluid with equation of state $p_\Lambda = - \rho_\Lambda$ (where $p_\Lambda$ and $\rho_\Lambda$ are the fluid pressure and energy density) or equation of state parameter $\omega = -1$. This ``standard'' $\Lambda$CDM model~\cite{peebles84}, where cold dark matter (CDM) is the second largest contributor to the cosmological energy budget after $\Lambda$, allows for a good fit to many curent observations.\footnote{The ``standard'' CDM model of structure formation might be inconsistent with some observations (see, e.g., \cite{peri}).} However, the needed $\Lambda$ energy density scale of order $10^{-3}$ eV is very small and difficult to reconcile with the much higher value that a naive application of quantum mechanics suggests.
Furthermore, as the density of nonrelativistic matter (the CDM and baryonic matter) decreases during the cosmological expansion while the $\Lambda$ energy density remains constant, these two energy densities are comparable in magnitude only for a relatively short time and it is puzzling why we happen to observe near this epoch (this is the so-called coincidence puzzle).

Motivated by the fact that instead of a cosmological constant, if the energy density responsible for the accelerated expansion --- the dark energy --- also decreased in time, the energy densities of nonrelativistic
matter and dark energy would be comparable for a somewhat longer period, it is of interest to consider models where the dark energy density decreases with time. A simple example is the XCDM parametrization of dark energy, where dark energy is modeled by a spatially homogeneous fluid with equation of state  $p_X = \omega_X \rho_X$ with $\omega_X < 0$. While the XCDM parametrization is widely used, it is well known that it is not a complete parametrization as it cannot describe spatial inhomogeneities (see, e.g. \cite{ratra91}).

The $\phi$CDM model \cite{rp}, in which dark energy is taken to be a scalar field $\phi$ with a potential  energy density $V(\phi)$ that results in a dark energy density that is close to spatially homogeneous but slowly decreasing in time, provides a complete description of dark energy, unlike the XCDM parametrization. See, e.g. \cite{frieman1995} for more recent discussions about this model.

Rather than introducing a new dark energy component, explanation of the accelerated cosmological expansion has been sought in modifications of GR, including the possible existence of extra dimensions (e.g. \cite{brane}) or by modifying GR by adding terms proportional to powers of the Ricci scalar $R$ to the Einstein-Hilbert Lagrangian (e.g. \cite{fr}).

Given the state of uncertainty that remains over whether $\Lambda$ may or may not provide a complete explanation for the observed cosmic acceleration, a very important way to improve our understanding of this phenomenon is to use cosmological observations to constrain mechanisms of cosmic acceleration. In this regard, it is worth emphasizing that using different observational techniques, as well as combinations of them, is particularly important for a more reliable determination of cosmological parameters, since different methods may rule out different regions of the parameter space and, therefore, may be complementary to each other.

In this paper we combine distance data with low and high-redshift time measurements to constrain accelerating cosmologies. Specifically, we use age measurements of 32 passively evolving galaxies~\cite{svj} (in the range $0.117 \leq z \leq 1.845$) to perform a statistical analysis of the age-$z$ test. In order to better constrain the parameter spaces of these models, we combine the age-$z$ data with a recent estimate of the ratio of the CMB acoustic scale $\ell_{A}$ and the baryonic acoustic oscillation (BAO) peak, the so-called CMB/BAO ratio~\cite{sollerman}. We show that for some models the age-$z$ plus CMB/BAO ratio data analysis results in restrictive joint constraints on these cosmologies. We discuss our results in terms of two information criteria, namely, the Akaike Information Criterion~\cite{aik} and the Bayesian Information Criterion~\cite{bic}.

The outline of our paper is as follows. In Sec.\ 2 we present the basic equations of, and main motivations for, the accelerating cosmological models we consider in our analyses. The age-redshift test and the joint analysis of this data with the CMB/BAO ratio data are outlined in Sec.\ 3. Our results are presented and discussed in Sec.\ 4. In Sec.\ 5 we summarize our main results.

\section{Nonstandard accelerating cosmologies}

As mentioned in the previous Section, theoretical models of cosmic acceleration can be broadly classed into two groups: those that introduce a new component to the composition of the Universe (this includes the ``standard'' $\Lambda$CDM case) and those that invoke a modification to the equations governing gravity.  In what follows, we briefly discuss some of the more popular nonstandard models\footnote{For other  alternative models, see \cite{other}.} and examine whether they are consistent with current age and distance  data. Unless specified otherwise, we assume vanishing spatial curvature.

\subsection{Quintessence scalar field}

For some potential energy density functions, $V(\phi)$, the $\phi$CDM model partially resolves two theoretical puzzles of the $\Lambda$CDM model. For example, with an inverse power potential $V(\phi) \propto \phi^{-\alpha}$ with $\alpha > 0$~\cite{rp} the solution of the coupled nonlinear equations of motion is an attractor for which the scalar field energy
density, $\rho_\phi$, decreases less rapidly than the dominant component of the energy density in the radiation- and matter-domianted epochs and so eventually comes to dominate, thus partially resolving the coincidence puzzle. Also, the slow decrease in time of the dark energy density in this model results in a very small dark energy density now because the Universe is old. Consequently, instead of a new fundamental energy density scale of order a meV as in the $\Lambda$CDM model, the $\phi$CDM model requires a higher fundamental energy density scale that is determined by when the phenomenological $\phi$CDM model first provides an appropriate description of cosmology.

With $V(\phi) = \kappa \phi^{-\alpha}$, where $\kappa$ is a constant, the dynamics of the $\phi$CDM model is obtained by solving the equations
\begin{equation}
H^2 = \frac{8\pi G}{3}(\rho_m + \rho_{\phi})\, ,
\end{equation}
\begin{equation}
\ddot{\phi} + 3H\dot{\phi} = {\kappa \alpha \over \phi^{\alpha + 1}}\, ,
\end{equation}
where $H$ is the Hubble parameter, $\rho_m$ is the pressureless matter (baryons and cold dark matter) energy density and a dot denotes a derivative with respect to time. The parameter $\alpha$ describes the steepness of the scalar field potential, with larger (smaller) values of $\alpha$ corresponding to faster (slower) evolution of the scalar field. In the limit $\alpha \rightarrow 0$ the spatially-flat $\phi$CDM model reduces to the spatially-flat $\Lambda$CDM case. The observational viability of this model has been extensively investigated in the literature and we refer the reader to \cite{rpmodel}.

\subsection{Chaplygin gas}

From the cosmological viewpoint, the main distinction between pressureless CDM and dark energy is that the former agglomerates at small length scales whereas the latter is a smooth component on these scales. Recently, the idea of a unified description for CDM and dark energy has received much attention (see, e.g., \cite{unified}). An interesting attempt in this direction was suggested in Ref.~\cite{kamen} and further developed in Ref.~\cite{bilic}. It uses to an exotic fluid, the so-called Chaplygin gas (Cg), whose equation of state is given by
\begin{equation} \label{cgeoS}
p_{Cg} = -\frac{A}{\rho_{Cg}^{\alpha}}\;.
\end{equation}
Actually, this equation for $\alpha \neq 1$ is a generalization of the original Cg equation of state and was originally proposed in \cite{bertolami} (see also~\cite{chaplygin}). In our analyses, we will consider both the original Cg ($\alpha = 1$) and its generalization (GCg).

Inserting Eq. (\ref{cgeoS}) into the energy conservation equation gives the expression for the GCg energy density
\begin{equation}
\rho_{Cg} = \rho_{Cg_{0}}\left[A_s + (1 - A_s)a^{-3(1 + \alpha)}\right]^{\frac{1}{1 +\alpha}},
\end{equation}
where $a(t)$ is the cosmological scale factor and $A_s = A/\rho_{Cg_{0}}^{1 + \alpha}$ is a quantity related to the sound speed for the generalized Chaplygin gas today (throughout this paper the subscript
``0'' denotes present-day quantities). As can be seen from the above equations, the Chaplygin gas interpolates between epochs dominated by non-relativistic matter [$\rho_{Cg}(a \rightarrow 0) \simeq {\rm{const.}}/a^{3}$] and by a negative-pressure time-independent dark ``energy'' [$\rho_{Cg}(a \rightarrow\infty) \simeq \sqrt{A}$].

\subsection{Coupled quintessence}

A phenomenological attempt at alleviating the cosmological constant problems mentioned earlier is based on allowing the dark matter and dark energy to interact.  For this class of models, the energy balance equation between a quintessence component $\phi$ and CDM particles must be of the form
\begin{equation}
\dot{\rho}_{cdm} + 3\frac{\dot{a}}{a}\rho_{cdm} = -\rho_{\phi} - 3\frac{\dot{a}}{a}\rho_{\phi}(1 + w_{\phi})\;.
\end{equation}
The dark matter-dark energy interaction implies that the energy density of the former component must dilute at a different rate compared to its standard evolution $\rho_{cdm} \propto a^{-3}$ (Alcaniz 2006). If we characterize this non-standard evolution by an arbitrary function $\epsilon(a)$ such that $\rho_{cdm} =  \rho_{cdm,0}a^{-3+\epsilon(a)}$ the energy density of the $\phi$ component can be written as (Costa 2010)
\begin{equation}\label{cq}
\rho_{\phi} =  \left[\rho_{\phi,0} + \rho_{cdm,0} \int_{a}^{1}\frac{[\epsilon(a) + \tilde{a} \epsilon^{'} ln \tilde{a}]}{\tilde{a}^{1 -3w_{\phi} - \epsilon(a)}} d\tilde{a} \right]a^{-3(1+w_{\phi})}\;.
\end{equation}
For $w_{\phi} = -1$ the above expression reduces to the vacuum decay model discussed by Costa \& Alcaniz (2010), whereas for a constant $\epsilon$ one recovers the coupled quintessence scenario studied by Jesus et al.\ (2008). Note also that the (incomplete) XCDM parametrization is recovered for $\epsilon = 0$.

For simplicity, we restrict our analyses to coupled quintessence models in which $\epsilon = \epsilon_0 \equiv \rm{const.}$ and $w _{\phi} = -1$, so that
\begin{equation} \label{decay}
 \rho_{\phi} =  \rho_{\phi,0} + \frac{\epsilon_0 \rho_{cdm,0}}{3 - \epsilon_0}a^{-3+\epsilon_0}\;,
\end{equation}
which is mathematically equivalent to the decaying vacuum scenarios proposed by Wang \& Meng (2005) and Alcaniz \& Lima (2005) (see Amendola 2000; Dalal et al.\ 2001; Wu et al.\ 2008; Carneiro et al.\ 2008, for other scenarios of coupled quintessence). Since we are assuming interaction only between the components of the dark sector, in our analyses of this model we fix the baryonic matter
fractional energy density contribution
at $\Omega_{b} = 0.0416$, as given by current CMB anistropy measurements (Komatsu et al.\ 2009) for Hubble constant $H_0 = 74.2$ km s$^{-1}$ Mpc$^{-1}$.

\subsection{DGP}

A different possible explanation of cosmic acceleration is a modification in gravity instead of an adjustment to the energy content of the Universe.\footnote{As mentioned earlier, in addition to the extra dimensional models we consider, another way of modifying gravity is by adding terms proportional to powers of the Ricci scalar $R$ to the Einstein-Hilbert Lagrangian, the so-called $f(R)$ gravity model. For some recent applications of $f(R)$ gravity in cosmology see Sotiriou (2009), de Felice \& Tsujikawa (2010), and references therein.} This motivates studies of cosmological models with extra spatial dimensions that might be able to explain the acceleration as well as provide a possible rationale for the  huge difference between the electroweak and Planck scales $m_{Pl}/m_{EW} \sim 10^{16}$ (Arkani-Hamed et al.\ 1998; Randall \& Sundrum 1999).

Many extra-dimensional models have been discussed. Here we consider the DGP model (Dvali et al.\ 2000), a self-accelera\-ting 5-dimensional model with a non-compact, infinite-volume extra dimension, whose gravitational dynamics is governed by a competition between a 4-dimensional Ricci scalar term induced on the brane  and an ordinary 5-dimensional Einstein-Hilbert action
(for a review of DGP phenomenology, see Lue 2006). The Friedmann equation derived from this gravitational action take the form
\begin{equation}
 H^2 + \frac{k}{a^2} = \left[\sqrt{\frac{\rho}{3M_{Pl}^2} + \frac{1}{4r_c^2} + \frac{1}{2r_c^2}}\right]^2\;,
\end{equation}
where $\rho$ is the energy density of the cosmic fluid, $k$ is the spatial
curvature, and $r_c = M^2_{Pl}/2M_5^3$ is the crossover scale defining the gravitational interaction among particles located on the brane. From the above equation we can also define the density parameter associated with the crossover radius, $\Omega_{rc} = 1/4r_c^2H_0^2$ (see also Deffayet et al.\ 2001; Alcaniz 2002; Guo et al.\ 2006). For this scenario we analyze both flat (flat DGP) and arbitrary-curvature cases (DGP).

\subsection{Cardassian expansion}

 Also based on extra dimensions, Freese \& Lewis (2002) proposed the Cardassian model in which the Universe is spatially flat, matter dominated, and currently undergoing accelerated expansion. Such a behavior is obtained from a modified Friedmann equation on our observable brane with $H^ 2 = g (\rho_m)$, where $g (\rho_m)$ is an arbitrary function of the matter energy density. Although completely different physically, the original Cardassian model with $g (\rho_m) = A^m + B^n$ (where $A$ and $B$ are functions of the matter density  $\rho_m$ and $n$ and $m$ are free parameters of the model) predicts the same observational effects as the XCDM parametrization for cosmological tests based on the evolution of the Hubble parameter with redshift.

Here, we consider the generalized Cardassian (GCard) model described by Wang et al.\ (2003) whose Friedmann equation is given by
\begin{equation} \label{hcard}
 H^2 = \frac{8\pi G \rho_m}{3}\left[1 + \left(\frac{\rho_{\rm{Card}}}{\rho_m}\right)^{q(1-n)}\right]^{1/q}\;,
\end{equation}
where $n$ and $q$ are two free parameters to be adjusted to fit the data and $\rho_{\rm{Card}} = \rho_{m,0}(1 + z_{\rm{Card}})^3$ is the energy density at which the two terms inside the bracket become equal. Note that, although matter dominated, GCard models are accelerating and  may still reconcile current indications for a spatially-flat Universe from CMB observations with clustering estimates that point consistently to $\Omega_m \simeq 0.3$. In these scenarios, this discrepancy is explained through a redefinition of the value of the critical density from Eq. (\ref{hcard}) (see, e.g.\ Freese \& Lewis 2002; Alcaniz et al.\ 2005 for discussions).

\section{Cosmological constraints}

\subsection{The age-redshift test}

The theoretical age-$z$ relation [$t(z_i)$] of an object at redshift $z_i$ can be written as (Sandage 1988)
\begin{eqnarray} \label{five}
t(z_i, \mathbf{p})  =  \int_{0}^{(1+z)^{-1}}{\frac{dx'}{x'{{\cal{H}}(x',\mathbf{p})}}}\;,
\end{eqnarray}
where $\mathbf{p}$ stands for the parameters of the cosmological model under consideration and ${\cal{H}}(x,\mathbf{p})$ is the normalized Hubble parameter. From the observational viewpoint, the total age of a given object (e.g., galaxies) at redshift $z$ is given by
$t^{\rm obs}(z_i)  =  t_G(z_i) + \tau$,
where where $t_G(z_i)$ is the estimated age of its oldest stellar population and $\tau$ is the incubation time or delay factor, which accounts for our ignorance about the amount of time since the beginning of structure formation in the Universe until the formation time of the object of interest.

To perform our age-$z$ analyses in the next section, we use age estimates of 32 old passive galaxies distributed over the redshift interval $0.117 \leq z \leq 1.845$ (Simon et al.\ 2005) as listed in Table 1 of Samushia et al.\ (2010), and assume a 12\% one standard deviation uncertainity on the age measurements (R.\ Jimenez, private communication 2007).
The total sample is composed of three sub-samples: 10 field early-type galaxies from Treu et al.\ (1999; 2001; 2002), whose ages were obtained by  using the SPEED models of Jimenez et al.\ (2004); 20 red galaxies from the publicly released Gemini Deep Deep Survey (GDDS), whose integrated light is fully dominated by evolved stars (Abraham et al. 2004; McCarthy et al. 2004); and the 2 radio galaxies LBDS 53W091 and LBDS 53W069 (Dunlop et al. 1996; Spinrad et al. 1997; Nolan et al. 2001). As discussed by McCarthy et al. (2004), the GDDS data seem to indicate that the most likely star formation history is that of a single burst of duration less than 0.1 Gyr, although in some cases the duration of the burst is consistent with 0 Gyr, which means that the galaxies have been evolving passively since their initial burst of star formation.

The likelihood function ${\cal{L}}\propto [\exp -\chi^2_{age}(z; \mathbf{p}, \tau)/2]$ is determined by a $\chi^2$ statistics
\begin{eqnarray} \label{chi2}
\chi_{age}^{2}(H_0, \tau, \mathbf{p})  =   \sum_{i=1}^{32}\frac{\left[t(z_i,\mathbf{p}) - t_G(z_i) - \tau\right]^{2}} {\sigma_{t_{G,i}}^{2}}\;, 
\end{eqnarray}
where $\sigma_{t_{G,i}}^{2}$ stands for the uncertainties on the age measurements of each galaxy in our sample. An important aspect in our analysis that is worth emphasizing  concerns the delay factor $\tau$. Note that in principle there must be a different $\tau_i$ for each object in the sample since galaxies form at different epochs. Here, however, since we do not know the formation redshift for each object, we assume a uniform delay factor $\tau$ that we treat as a ``nuisance'' parameter and marginalize over it. Similarly, we also marginalize over the present value of the Hubble parameter $H_0$.\footnote{A variant of this test uses both measurements of the Hubble parameter as a function of redshift (see, e.g.\ Samushia \& Ratra 2006; Samushia et al.\ 2007; Dev et al.\ 2008; Gong et al.\ 2008; Fernandez-Martinez \& Verde 2008; Wei 2009; Yang \& Zhang 2009) and lookback time measurements built from estimates of the total age of the Universe (Capozziello et al.\ (2004), Pires et al.\ (2006), Dantas et al. (2007, 2009), Samushia et al.\ (2010) and Samushia (2009).} Note also that, in order to avoid double counting of information with CMB data discussed in the next section, we have not included any prior on the age of the Universe from CMB data, as usually done in lookback time tests.

\begin{figure}[t]
\centerline{\psfig{figure=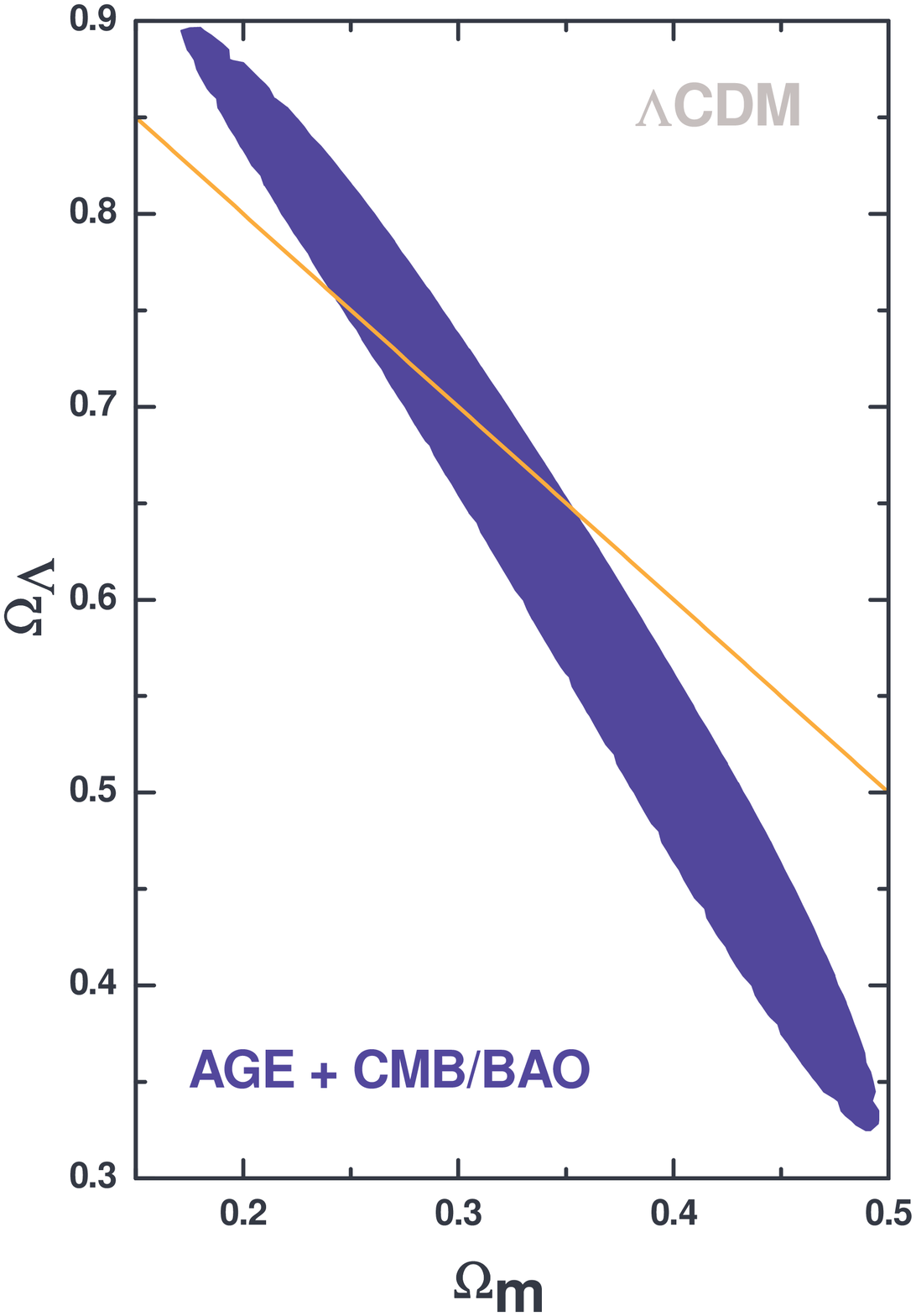,width=1.8truein,height=2.7truein,angle=0}
\psfig{figure=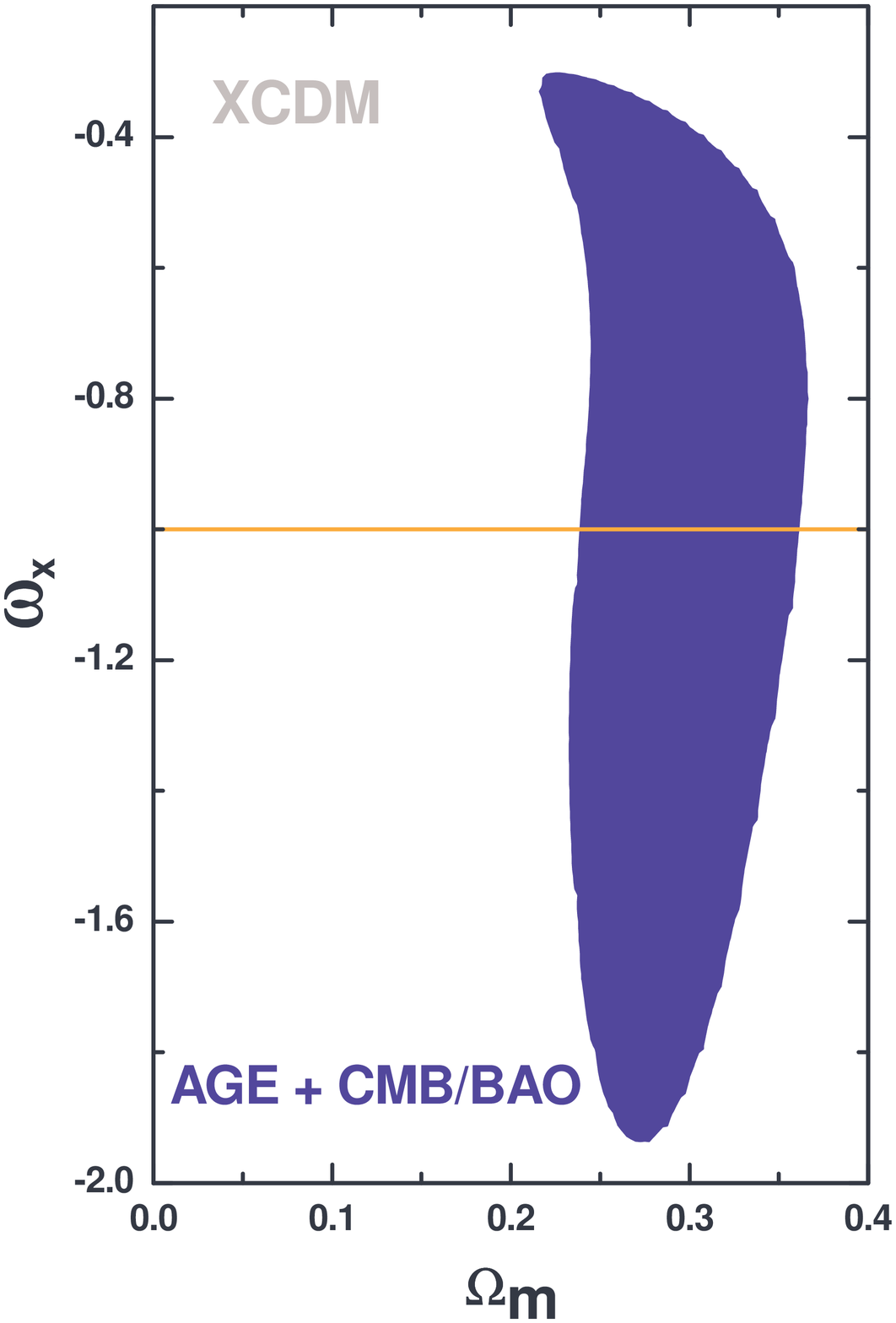,width=1.8truein,height=2.7truein,angle=0}}
\caption{Contours of $\Delta \chi^2_T = 6.17$ (2$\sigma$) for the $\Lambda$CDM model (left panel) and the spatially-flat XCDM parametrization (right panel).  The contours correspond to the joint analysis involving age-$z$ plus CMB/BAO data. In the left panel ($\Lambda$CDM) the thin orange diagonal line demarcates spatially-flat models while in the left panel (XCDM) the thin orange horizontal line indicates models with a time-independent cosmological constant.}
\label{fig1}
\end{figure}

\subsection{CMB/BAO ratio}

The two major inputs involving acoustic oscillations come from the CMB and baryon oscillations data. Although the CMB shift parameter ${\cal{R}}$ and the BAO parameter ${\cal{A}}$ have been commonly used to constrain nonstandard models, their use may not always be entirely appropriate because the values used were obtained in the context of an extended XCDM parametrization, which can be considered a good approximation only for some classes of dark energy models (see, e.g.\ Doran et al.\ 2007).\footnote{For the BAO parameter ${\cal{A}}$, for instance, it is implicitly assumed that the evolution of matter density perturbations during the matter-dominated era must be similar to the $\Lambda$CDM case and also that the comoving distance to the horizon at the time of equilibrium between matter and radiation energy
densities must scale with $(\Omega_mH_0^2)^{-1}$.}

\begin{figure}[t]
\centerline{\psfig{figure=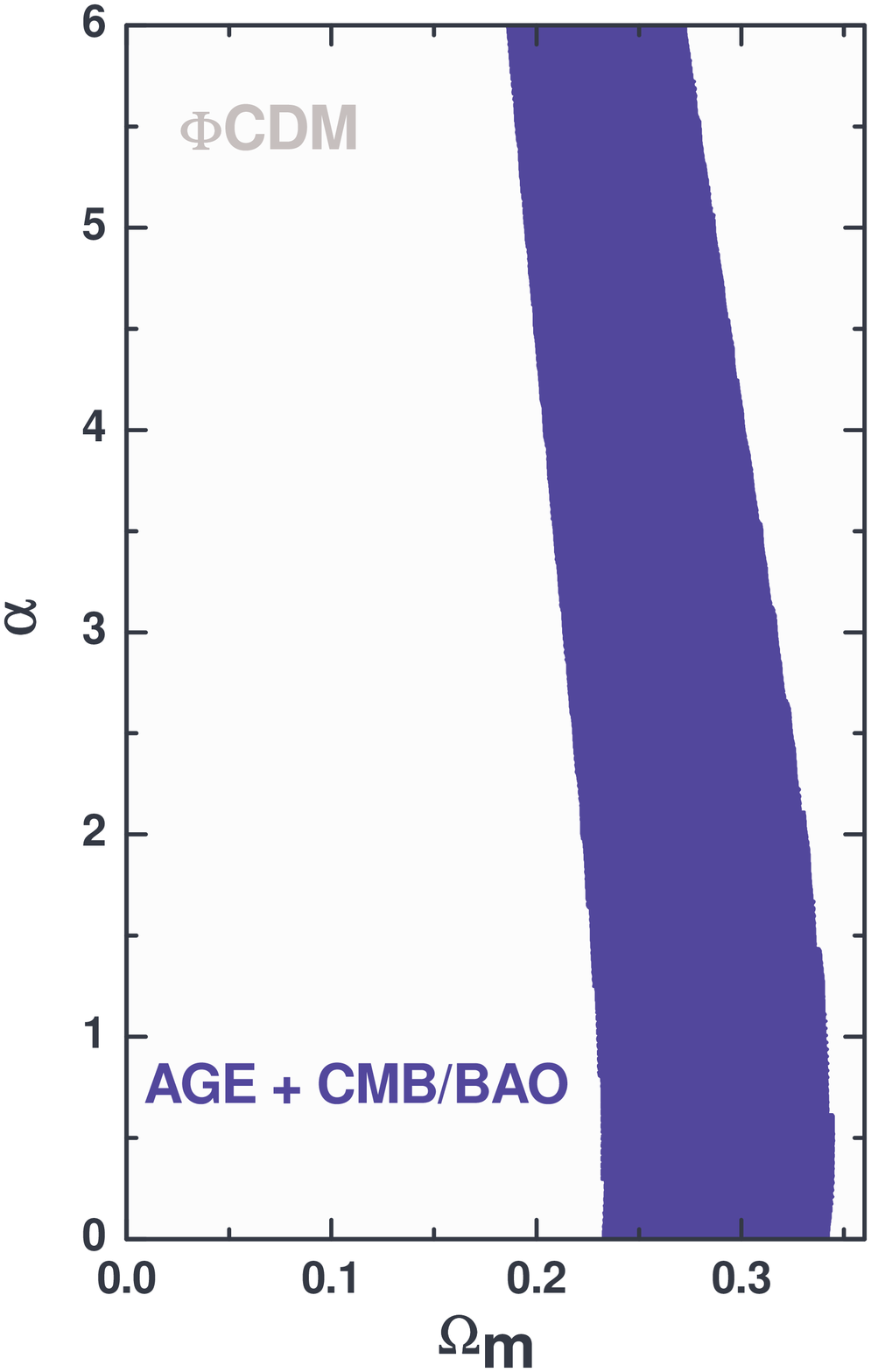,width=1.8truein,height=2.7truein,angle=0}
\psfig{figure=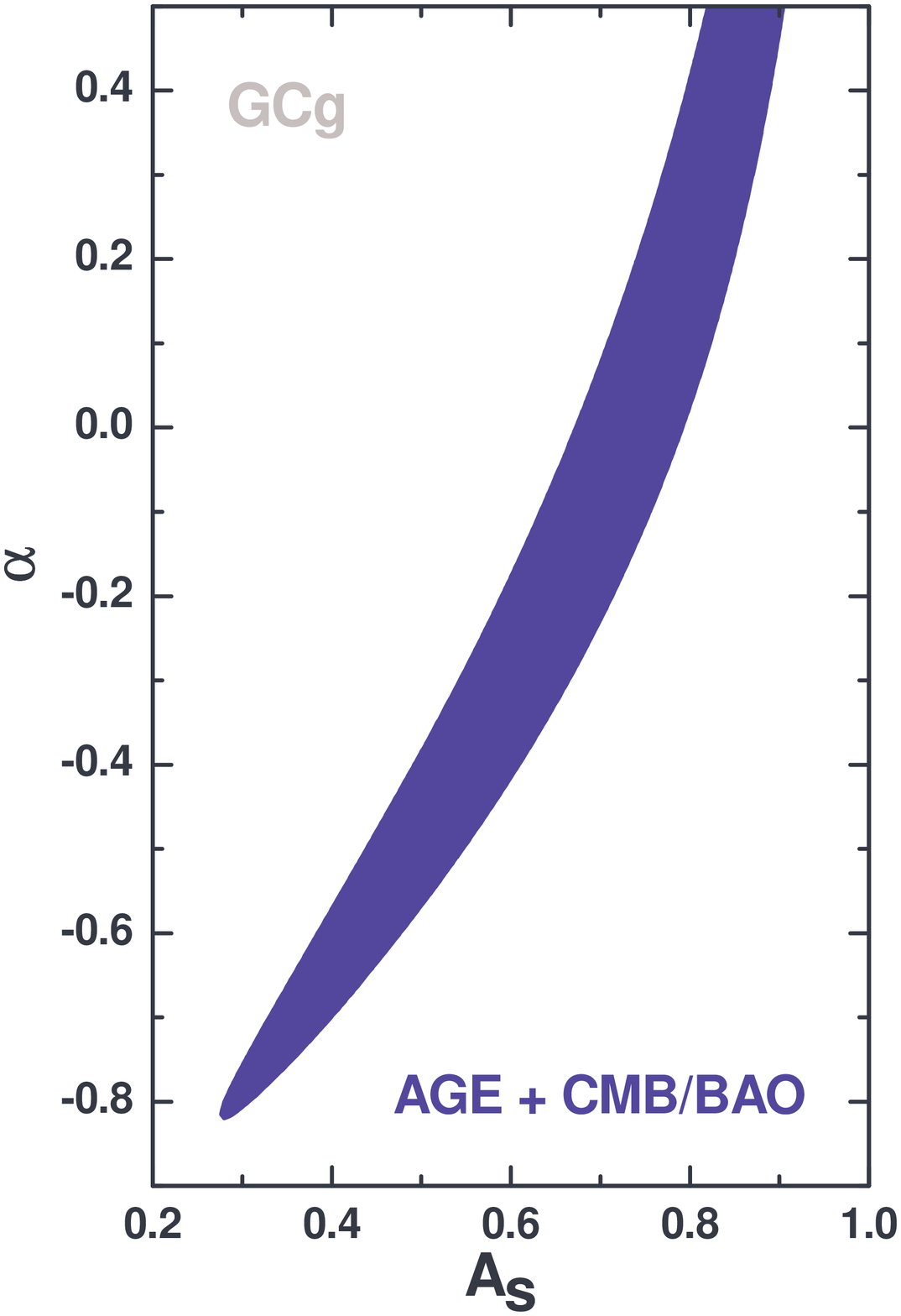,width=1.8truein,height=2.7truein,angle=0}}
\caption{Contours of $\chi^2_T$ corresponding to the joint analysis discussed in the text for two of the accelerating cosmologies. Left panel shows the spatially-flat $\phi$CDM model with $V(\phi) \propto \phi^{-\alpha}$ while the right panel is for the Generalized Chaplygin gas case.  The contours are drawn for $\Delta \chi^2 = 6.17$ (2$\sigma$). For $\alpha = 0$ the spatially-flat $\phi$CDM model reduces to the spatially-flat $\Lambda$CDM case.}
\end{figure}

\begin{figure*}
\centerline{\psfig{figure=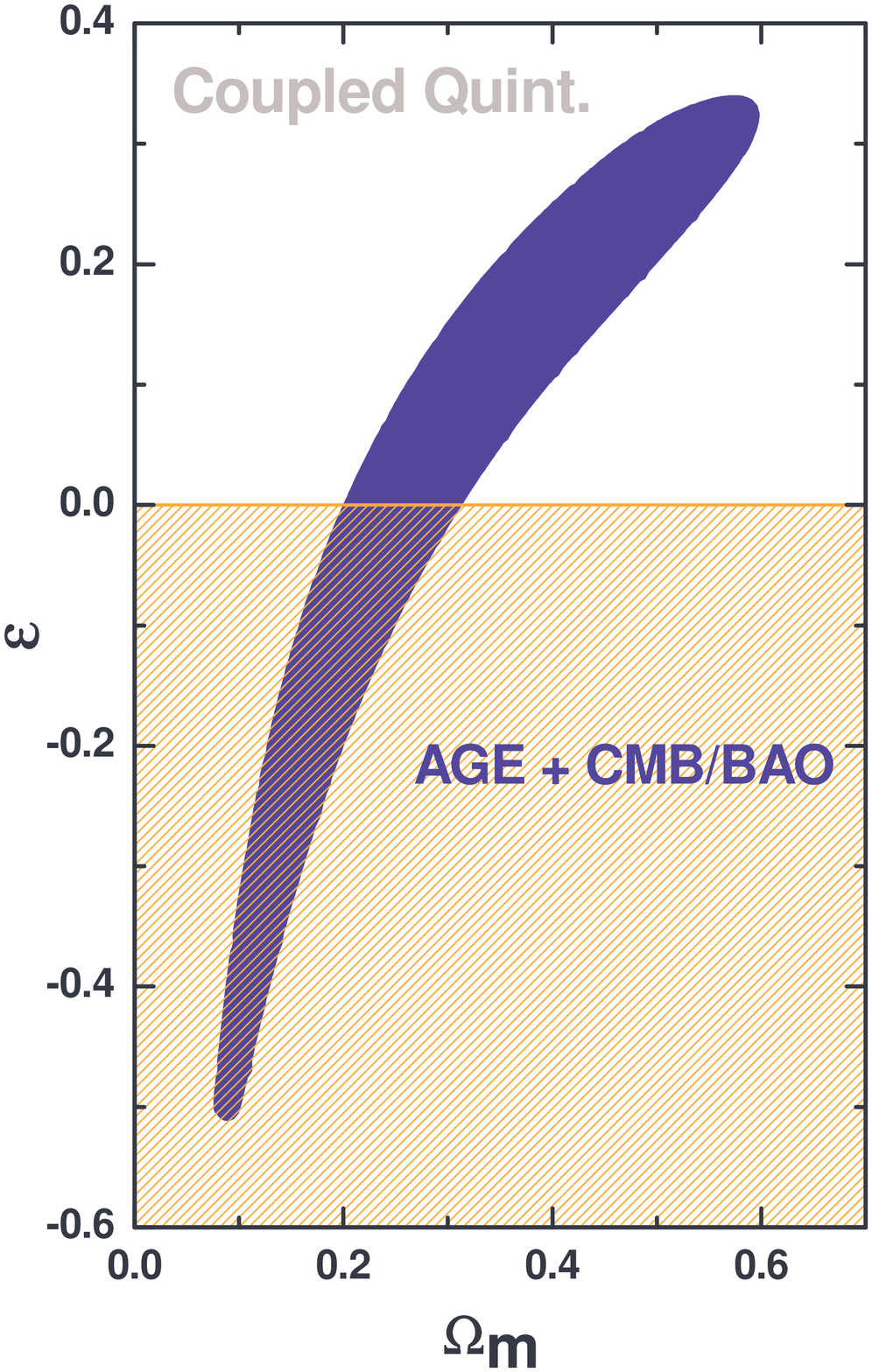,width=2.4truein,height=3.0truein,angle=0}
\psfig{figure=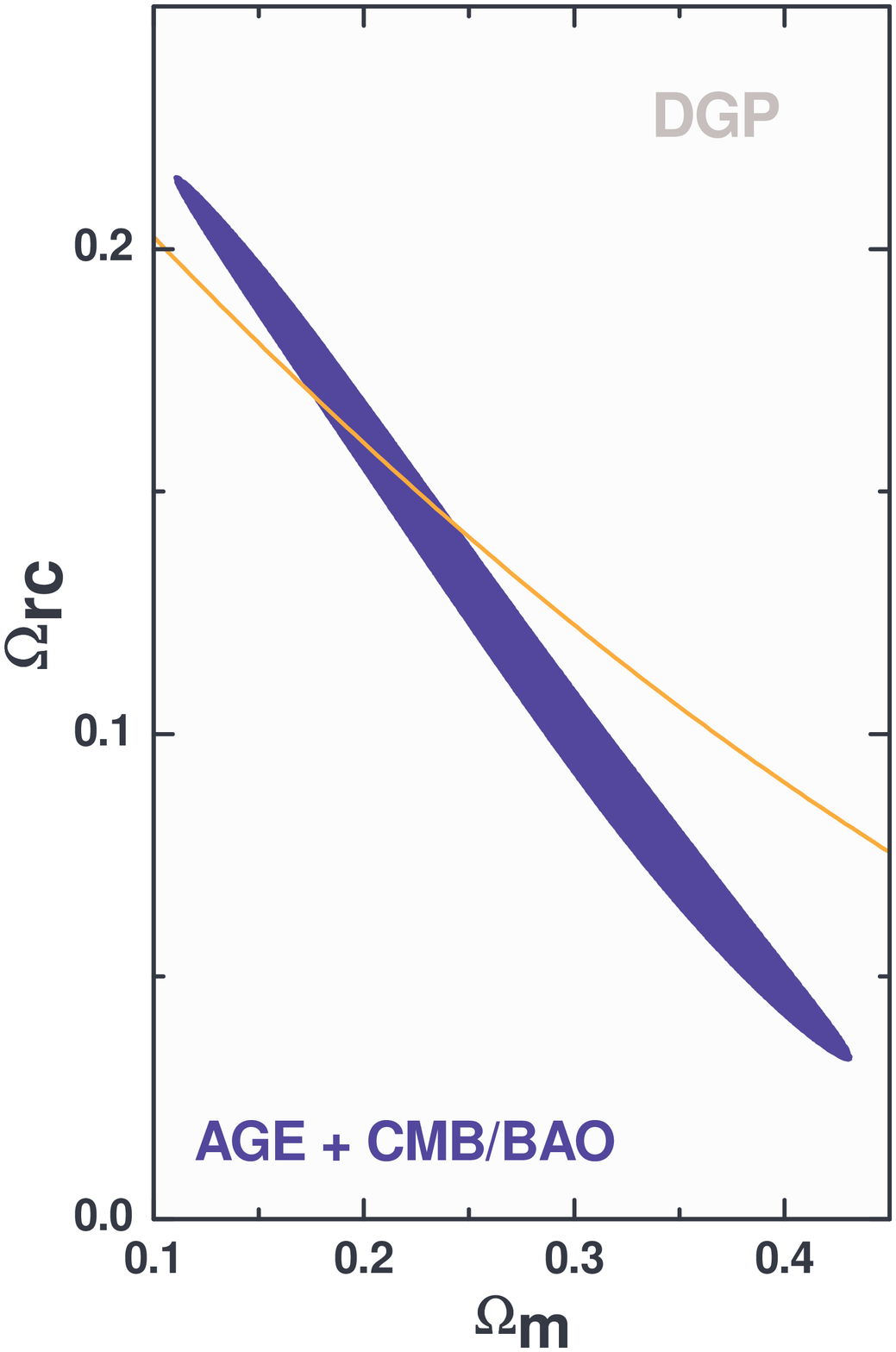,width=2.4truein,height=3.0truein,angle=0}
\psfig{figure=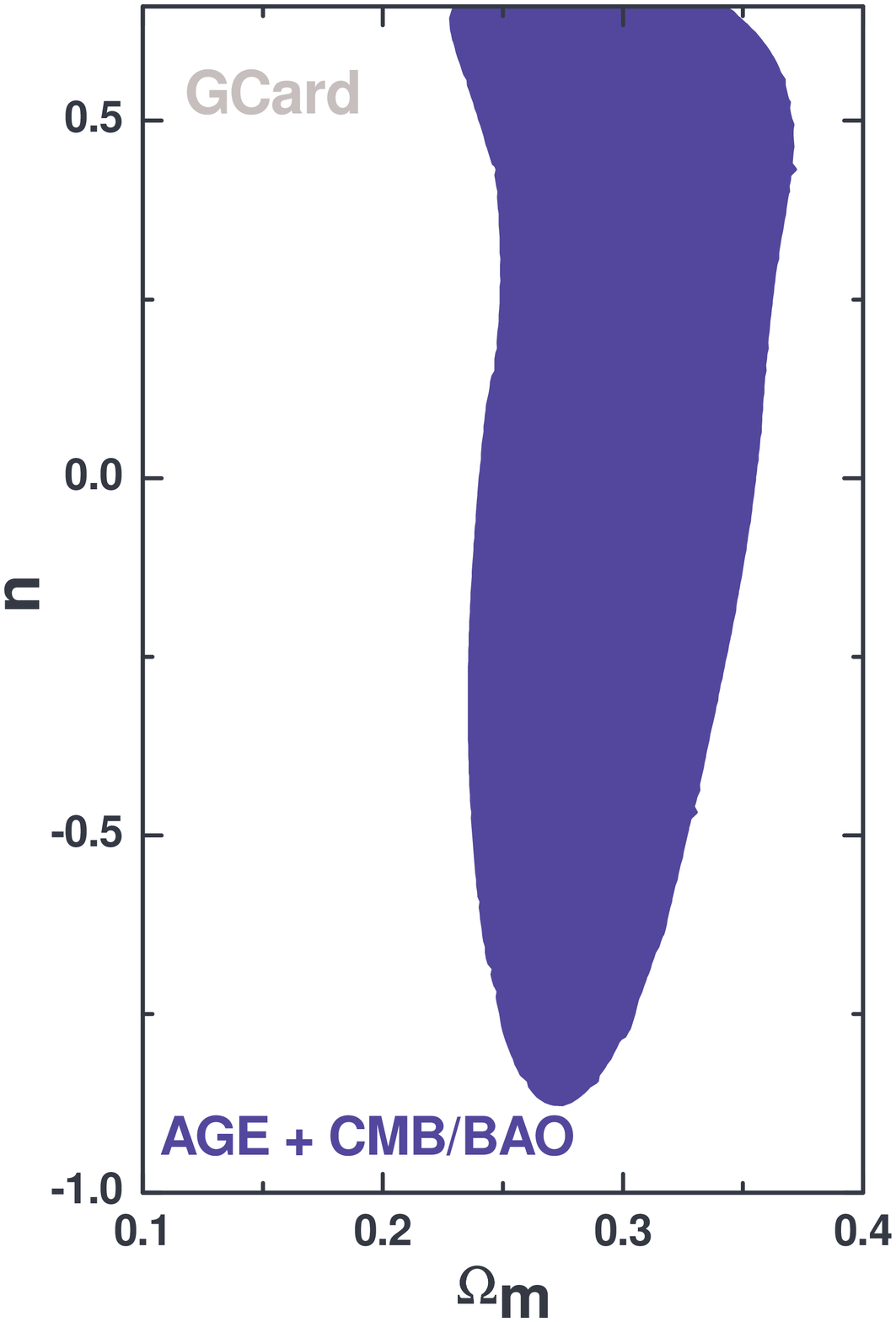,width=2.4truein,height=3.0truein,angle=0}}
\caption{Contours of constant $\Delta \chi^2_T = 6.17$ for the spatially-flat coupled  quintessence (left panel),  DGP (middle panel), and generalized Cardassian  (right panel) models. The shadowed area in the left panel (coupled quintessence) corresponds to the thermodynamical constraint on the interaction parameter ($\epsilon_0 \geq 0$) discussed by Alcaniz \& Lima (2005). The thin orange line in the middle panel (DGP) demarcates spatially-flat models.}
\end{figure*}

Here, we follow Sollerman et al.\ (2009) and use a more model-independent constraint derived from the product of the CMB acoustic scale $\ell_{A} = \pi d_A (z_*)/r_s(z_*)$ and the measurement of the ratio of the sound horizon scale at the drag epoch and the BAO dilation scale, $r_s(z_d )/D_V(z_{\rm{BAO}})$ [$d_A (z_*)$ is the comoving angular-diameter distance to recombination and $r_s(z_*)$ is the comoving sound horizon at decoupling]. By combining the ratio $r_s (z_d = 1020)/r_s (z_*=1090) = 1.044 \pm 0.019$ (Komatsu et al.\ 2009) with the measurements of $r_s(z_d)/D_V(z_{\rm{BAO}})$ at $z_{\rm{BAO}} = 0.20$ and 0.35 from Percival et al.\ (2009), Sollerman et al.\ (2009) found (with one standard deviation error bars)
$$
d_A (z_*)/D_V (0.2) = 17.55 \pm0.65
$$
$$
d_A (z_*)/D_V (0.35) = 10.10 \pm 0.38\;,
$$
which we use in our analyses along with the LT-$z$ data discussed earlier. We account for the correlations in the measurements by following Sollerman et al.\ (2009).

\section{Results}


Figures 1, 2, and 3 show some of the main results of our analyses. In these figures, we show $2\sigma$ contours for the the joint analysis involving age-$z$ plus CMB/BAO data. Note that, although not providing very restrictive constraints in all the cases analized, these time and distance cosmological data can place tighter bounds on some cosmological parameter spaces. In particular, for the $\Lambda$CDM (Fig.\ 1 left panel) and DGP (Fig.\ 3 center panel) cases the $\chi^2$ contours for age-$z$ and CMB/BAO data are roughly orthogonal, which results in tight cosmological constraints when used together (see, e.g., Pires et al.\ 2006; Samushia et al.\ 2010).

The shadowed area in the left panel of Fig.\ 3 for the coupled quintessence case corresponds to the thermodynamical constraint on the interaction parameter ($\epsilon_0 \geq 0$) discussed by  Alcaniz \& Lima (2005). In agreement with other recent analyses (Bertolami et al.\ 2007; Costa et al.\ 2009), we note that half of the $2\sigma$ interval for $\epsilon_0$ lies in the negative region of the plot, which means that an energy flow from dark energy to dark matter is allowed from this combination of data (see Eq. (\ref{decay})).

In order to select the best-fit models, we use two information criteria, namely, the Bayesian information criterion (BIC) and the Akaike information
criterion (AIC):
\begin{equation}
 {\rm{BIC}} = -2\ln{\cal{L}} + k\ln N\;,
\end{equation}
\begin{equation}
 {\rm{AIC}} = -2\ln{\cal{L}} + 2k\;,
\end{equation}
where ${\cal{L}}$ is the maximum likelihood, $k$ is the number of model parameters, and $N$ is the number of data points used in the fit. We refer the reader to Liddle (2004), God\l{}owski \& Szydy\l{}owski (2005), Biesiada (2007), and Davis et al.\ (2007) for reviews on the background for the use of these ICs. ICs provide an interesting way to obtain a relative ranking of the observational viability of different candidate models. Thus, the important quantities in these analyses are the differences $\Delta {\rm{AIC}}_i =  {\rm{AIC}}_i - {\rm{AIC}}_{{\rm min}}$ and $\Delta {\rm{BIC}}_i =  {\rm{BIC}}_i - {\rm{BIC}}_{{\rm min}}$ calculated over the whole set of scenarios $i = 1, ..., N$ with the best-fit model being the one that minimizes the {AIC} and {BIC} factors. Here, we follow
Liddle (2004) and adopt the criterion that a difference of 2 is considered positive evidence against the model with the higher BIC, while a $\Delta {\rm{BIC}}$ of 6 is considered strong evidence against the model (we also adopt similar rules for the  AIC). Note also that, since in our analysis $N/k < 40$, we use the version of AIC corrected for small samples ${\rm{AIC}_c} = {\rm{AIC}} +  2k(k + 1)/(N - k - 1)$ (Sugiura 1978; Burnham \& Anderson 2002, 2004; Liddle 2004).

\begin{table}[t]
\begin{center}
\caption{Summary of the age-$z$ IC Results}
\begin{tabular}{rrll}
\hline  \hline \\
\multicolumn{1}{c}{Model}&
\multicolumn{1}{c}{$k$}&
\multicolumn{1}{c}{$\Delta$AIC}&
\multicolumn{1}{c}{$\Delta$BIC} \\ \hline  \\ 
Flat XCDM         & 2 &(5) 2.78 & 3.97 (4)\\ 
Flat $\Lambda$CDM & 1 &(3) 0.94& 0.94  (2)\\ 
$\Lambda$CDM      & 2 &(8) 2.85& 4.03  (7)\\ 
Flat DGP          & 1  &(1) 0.00 & 0.00  (1)\\ 
DGP               & 2  &(2) 0.93& 4.11 (8)\\ 
Flat $\phi$CDM    & 2 &(7)  2.80 & 3.99 (6)\\ 
Chaplygin         & 1  &(4) 1.91& 1.90  (3)\\ 
Gen.\ Chaplygin   & 2  &(6) 2.79& 3.98  (5)\\ 
Gen.\ Cardasssian & 3  &(9) 4.58& 7.40  (9)\\ 
Coupled Quint.\   & 2  &(6) 2.79& 3.98  (5)\\ 
\hline  \hline
\end{tabular}
\end{center}
\end{table}

\begin{table} [t]
\begin{center}
\caption{Summary of the age-$z$ and CMB/BAO IC Results}
\begin{tabular}{rrll}
\hline  \hline \\
\multicolumn{1}{c}{Model}&
\multicolumn{1}{c}{$k$}&
\multicolumn{1}{c}{$\Delta$AIC}&
\multicolumn{1}{c}{$\Delta$BIC}\\   \hline  \\ 
Flat XCDM         & 2 &(9)  1.87& 3.14 (9)\\ 
Flat $\Lambda$CDM & 1 &(1)  0.00& 0.00 (1)\\ 
$\Lambda$CDM      & 2 &(7)  1.58& 2.85 (6)\\ 
Flat DGP          & 1  &(5) 1.45& 3.00 (7)\\ 
DGP               & 2  &(4)  1.40& 2.67 (3)\\ 
Flat $\phi$CDM    & 2 &(6) 1.52  &  2.80 (5)\\ 
Chaplygin         & 1  &(3)  1.33& 1.54 (2)\\ 
Gen.\ Chaplygin   & 2  &(8)  1.80& 3.07 (8)\\ 
Gen.\ Cardasssian & 3  &(10)  4.23& 6.54 (10)\\ 
Coupled Quint.\   & 2  &(2)  1.04& 2.69 (4)\\ 
\hline  \hline
\end{tabular}
\end{center}
\end{table}

Table 1 shows a summary of the IC results for the age-$z$ test. The best-fit model from both ICs is the spatially-flat DGP model, in contrast to earlier analyses, except for the result shown in Table 2 of Sollerman et al.\ (2009) based on the SDSS SNe Ia sample (MLCS-fit) and the CMB/BAO ratio data discussed above. Following the rule for the BIC discussed earlier, we note that there is mostly only positive evidence in favor of the best-fit scenario relative to most of the models considered ($\Delta$BIC $\simeq 2$). The exceptions are the DGP model with arbitrary spatial curvature ($\Delta$BIC = 4.11) and the generalized Cardassian expansion scenario ($\Delta$BIC = 7.40), where the evidence in favor the best-fit model is considerably stronger.

These results are considerably modified when the CMB/BAO data are added to the analysis (Table 2). The best-fit model now is the flat $\Lambda$CDM scenario, and there is now strong evidence against only the generalized Cardassian scenario ($\Delta$BIC = 6.54). It is also worth mentioning that five (Coupled Quint., DGP, Chaplygin gas, spatially-flat $\phi$CDM, generalized Chaplygin gas) out of the seven non-standard cosmologies considered in this paper cannot be ruled out by this combination of data. The results found in our joint analyses reflect the complementarity between age-$z$ and CMB/BAO measurements and confirm our initial arguments for a joint analysis involving time and distance-based cosmological observables.

\section{Conclusion}

We have investigated the observational viability of some non-standard accelerating cosmologies in light of current observational data. In contrast to most recent analyses, which are essentially based on distance measurements to a particular class of objects or physical rulers, we have used complementary measurements involving time (age-$z$) and distance (CMB/BAO) data.

We have shown that some alternative models not only survive the investigation but in some cases also provide a better fit to the data than does the current standard $\Lambda$CDM cosmological scenario (see Tables 1 and 2). These results are in good agreement with recent analyses using SNe Ia plus CMB/BAO data (see, e.g., Sollerman et al.\ 2009) and reinforces the need for investigations of alternative mechanisms of cosmic acceleration that rely on different kinds of observations.

While current age-$z$ data alone are not able to strongly discriminate between different cosmological models, we expect a new higher quality set of high-$z$ age measurements to soon become avaliable (R.\ Jimenez, private communication, 2009). We anticipate that this new data will prove very useful in narrowing down the range of viable cosmological models.

\acknowledgments

The authors acknowledge financial support from CNPq grant 481784/2008-0 and DOE grant DE-FG03-99EP41093. DM is supported by the SNSF (SCOPES grant No 128040).

\end{document}